\documentclass[
reprint,
superscriptaddress,
amsmath,amssymb,
aps,
]{revtex4-2}

\usepackage{color}
\usepackage{graphicx}
\usepackage[FIGTOPCAP]{subfigure}
\usepackage{dcolumn}
\usepackage{dcolumn}
\usepackage{bm,braket}
\usepackage{ulem}

\begin {document}

\title{
    Unexpected Effects of Disorder on Current Fluctuations in the Symmetric Simple Exclusion Process
    }

\author{Issei Sakai}
\email{6224701@ed.tus.ac.jp}
\affiliation{%
    Department of Physics and Astronomy, Tokyo University of Science, Noda, Chiba 278-8510, Japan
}%

\author{Takuma Akimoto}
\email{takuma@rs.tus.ac.jp}
\affiliation{%
    Department of Physics and Astronomy, Tokyo University of Science, Noda, Chiba 278-8510, Japan
}%



\date{\today}

\begin{abstract}
    We explore how the disorder impacts the current fluctuations in the symmetric simple exclusion process (SSEP) within a heterogeneous environment.
    First, we analyze the SSEP with a defect site under the periodic boundary conditions.
    We derive the exact expression for the second moment of the current
    and observe deviations from that of the homogeneous system.
    Notably, the second moment of the current shows asymmetric density dependence around a density of $1/2$
    and surpassing that of the homogeneous system in the low-density region.
    Furthermore, based on the finding from the SSEP with a defect site, we present an approximate derivation of the second moment of the current
    in the SSEP on a quenched random energy landscape using a partial-mean-field approach.
    The second moment of the current is heavily influenced by the energy landscape,
    revealing unique effects arising from the interplay between the heterogeneous environment and the many-body system.
    Our findings provide valuable insights that can be applied to control current fluctuations in systems involving the interactions of many particles, such as biological transport.
\end{abstract}

\maketitle


\section{Introduction}
Single file diffusion, where particles diffuse in a narrow channel and cannot pass through each other, is a pedagogical model for interacting many-body systems.
It is well known that, on the infinite line, the mean square displacement (MSD) of the tracer particle
is given by $\Braket{\delta x(t)^2}\propto t^{1/2}$ in the long time limit \cite{Harris:1965aa},
where $\delta x(t)$ is the displacement of the tracer particle and $\Braket{\cdot}$ is the ensemble average.
Various transport phenomena in narrow channels show single file diffusion, for example,
molecular diffusion in zeolites \cite{PhysRevLett.76.2762,Kukla:1996aa},
DNA-binding proteins \cite{Li:2009aa},
transport of confined colloidal particles \cite{doi:10.1126/science.287.5453.625,PhysRevLett.93.026001},
and transport of water molecules in carbon nanotubes \cite{PhysRevLett.89.064503}.

The symmetric simple exclusion process (SSEP) is a fundamental model of single-file diffusion \cite{SPITZER1970246},
where particles with hard-core interactions move on a one-dimensional lattice.
With its simplicity, this model has led to the derivation of mathematically rigorous results.
On the infinite lattice, the large deviation function of the integrated current through the origin has been computed
independently by Bethe ansatz \cite{Derrida:2009aa} and the macroscopic fluctuation theory \cite{PhysRevLett.129.040601}.
Additionally, the large deviation function of the tracer position has been derived \cite{PhysRevLett.118.160601,Imamura:2021aa}.
The generalized density profile, which is the correlation between the observation (for example, the tracer position, the integrated current through the origin,
and the generalized current) and the density, has also been derived \cite{PhysRevLett.127.220601,Grabsch:aa,PhysRevE.107.044131}.
On the semi-infinite lattice connected to a reservoir, the cumulant generating function of the integrated current has been derived using the macroscopic fluctuation theory \cite{PhysRevLett.133.117102}.
Under the periodic boundary condition, the cumulants of the integrated current have been obtained \cite{PhysRevE.78.021122}.
In particular, the second moment of the total integrated current $Q(t)$ in time $t$
is derived as a function of the particle density $\rho$:
\begin{equation}
    \lim_{t\rightarrow\infty}\frac{d\Braket{Q(t)^2}}{dt}
    =\frac{L\rho(1-\rho)}{\tau},
    \label{current fluctuations for homogeneous system}
\end{equation}
where $\tau$ is the mean waiting time, and $L$ is the system size \cite{PhysRevE.78.021122}.

Single-particle diffusion in a heterogeneous environment shows various anomalous behaviors.
Such environments are represented by random energy landscapes.
The random energy landscape can be categorized into two types: an annealed energy landscape, which changes with time, and a quenched energy landscape, which remains static.
The random walk on an annealed energy landscape is often modeled by the continuous-time random walk.
In this model, waiting times are independent and identically distributed (IID) random variables, and the MSD exhibits anomalous diffusion when the mean waiting time diverges \cite{METZLER20001}.
The random walk on a quenched energy landscape is called the quenched trap model \cite{BOUCHAUD1990127}.
In the infinite system, the MSD shows anomalous diffusion when the mean waiting time diverges \cite{BOUCHAUD1990127}.
In the finite system, observables such as the drift, the diffusion coefficient, and the first passage time exhibit sample-to-sample fluctuations,
resulting in non-self-averaging behaviors \cite{AkimotoBarkaiSaito,*AkimotoBarkaiSaito2018,LuoYi,AkimotoSaito2019,AkimotoSaito2020}.

The influence of heterogeneity on the single-file diffusion has garnered significant attention in the field of statistical mechanics \cite{PhysRevE.108.054125,Metzler:2014aa,Sanders_2014,PhysRevLett.80.85,PhysRevLett.102.190602}.
Recently, the single file diffusion in the system with periodically varying potential and diffusivity has been studied,
and the MSD of the tracer particle has been obtained \cite{PhysRevE.108.054125}.
Specifically, the formula of the MSD is the same as that for the normal single-file diffusion.
However, the diffusion coefficient is replaced with the effective diffusion coefficient for a single particle diffusing in the system with periodically varying potential and diffusivity.
For the SSEP on an annealed energy landscape, when the mean or variance of waiting time diverges,
the MSD is $\Braket{\delta x(t)^2}\propto \log^{1/2}t$ or $\Braket{\delta x(t)^2}\propto t^{\gamma}$ with $0<\gamma<1/2$,
respectively \cite{Metzler:2014aa,Sanders_2014},
i.e., the diffusion in the SSEP on an annealed energy landscape is slower than in the normal SSEP.
The impacts of disorder in the asymmetric simple exclusion process (ASEP), where particles in the SSEP are influenced by an external field like an electric field, have been a subject of research for many years
\cite{PhysRevLett.78.3039,TripathyBarma,HarrisStinchcombe,Nossan_2013,10.1214/14-BJPS277,BanerjeeBasu,PhysRevE.107.L052103,*PhysRevE.107.054131,Enaud_2004,PhysRevLett.94.010601,JuhaszSantenIgloi,ConcannonBlythe,Goswami_2022}.
Under the periodic boundary condition, while, in the ASEP in the homogeneous environment, the relation between current $J$ and particle density $\rho$ is given by $J=\rho(1-\rho)$, in the ASEP on a quenched random energy landscape, it becomes flat around $\rho=1/2$ \cite{PhysRevLett.78.3039,TripathyBarma,HarrisStinchcombe,Nossan_2013,10.1214/14-BJPS277,BanerjeeBasu,PhysRevE.107.L052103,PhysRevE.107.054131}.
Furthermore, the current and the diffusion coefficient exhibit sample-to-sample fluctuations \cite{PhysRevE.107.L052103,*PhysRevE.107.054131}.
While several effects resulting from the interplay between a heterogeneous environment and a many-body system have been uncovered,
the study of the SSEP on a quenched random energy landscape remains unexplored.
Such heterogeneity is prevalent in various diffusion phenomena, such as the movement of
proteins on DNA~\cite{GraneliGreeneRobertsonYeykal,AustinCoxWang} and
water transportation in aquaporin~\cite{AkimotoHiraoYamamotoYasuiYasuoka} and OmpF~\cite{10.1063/1.2761897}.
Therefore, understanding current fluctuations in the SSEP on a quenched random energy landscape is crucial.

In this paper, we explore the impact of the disorder on the current fluctuation in the SSEP within a heterogeneous environment.
We specifically investigate the SSEP with a defect site, which serves as a basic representation of a heterogeneous environment.
By deriving the second moment of the total integrated current and comparing it to that of a homogeneous system,
we observe asymmetric density dependence in the second moment of the total integrated current around $\rho=1/2$
due to the heterogeneous environment.
Additionally, we extend our analysis to the SSEP on a quenched random energy landscape, where we discover a strong dependence of the second moment of the total integrated current on the energy landscape.
Remarkably, in a system where sample-to-sample fluctuations in single-particle diffusion are absent,
the introduction of many-body effects can give rise to sample-to-sample fluctuations.

Our paper is organized as follows.
In Sec.~\ref{sec: model}, we formulate the SSEP with a defect site and derive the site density and the two-point correlation.
In Sec.~\ref{sec: result}, we derive the second moment of the integrated current and compare the theoretical result with the numerical result.
In Sec.~\ref{sec: Application}, we apply the result of the SSEP with a defect site to the SSEP on a quenched random energy landscape.
Sections~\ref{sec: discussion} and \ref{sec: Conclusion} are devoted to the discussion and the conclusion, respectively.
In Appendix~\ref{sec: appendix correlation},
we prove that the correlation between the integrated current and the occupation number does not depend on the time.

\section{Model}\label{sec: model}
We study the SSEP with a defect site to unravel the effects of heterogeneity of the environment.
The system consists of $N$ particles on the lattice of $L$ sites with periodic boundary conditions, where each site can accommodate at most one particle.
The defect site has a longer mean waiting time compared to other sites, inhibiting particle movement.
The mean waiting times at site $i$ are defined as follows:
\begin{equation}
    \tau_i=
    \begin{cases}
        \mu\ (>1) & (i=1)\\
        1 &(i\neq 1).
    \end{cases}
\end{equation}
The waiting times are IID random variables following an exponential distribution,
$\psi^{(i)}(\tau)=\tau_i^{-1}\exp(-\tau/\tau_i)$.

We explicitly describe the dynamics of particles.
Initially, the particles are in an equilibrium ensemble of configurations.
The equilibrium ensemble is given by the configuration after particles are randomly placed and diffuse for a long time.
In infinite systems, the effects of the initial ensemble are everlasting \cite{PhysRevE.88.032107}.
At time $t=0$, each particle $i$ is assigned the waiting time $t_i$ from $\psi^{(x_i)}(t_i)$,
where $x_i$ is the site number of particle $i$.
Once the waiting time $t_i$ has elapsed, particle $i$ attempts to jump to the right site with a probability of $1/2$
and to the left site with a probability of $1/2$.
The jump is successful only if the target site is unoccupied.
Following a successful or unsuccessful jump attempt, 
particle $i$ is given a new waiting time based on either $\psi^{(x_i\pm1)}(t_i)$ or $\psi^{(x_i)}(t_i)$, respectively.

A configuration $C=\{\eta_i\}_{i=1,\dots,L}$ is represented by $L$ occupation numbers $\eta_i$,
where $\eta_i=1$ if site $i$ is occupied and $\eta_i=0$ otherwise.
Let $P_t(C)$ be the probability of finding the configuration $C$ at time $t$.
The master equation for $P_t(C)$ can be written as
\begin{equation}
    \frac{dP_t(C)}{dt}=\sum_{C'}\left[W(C',C)P_t(C')-W(C,C')P_t(C')\right],
    \label{master equation}
\end{equation}
where $W(C',C)$ is the rate at the configuration transition from $C'$ to $C$.
For $t\rightarrow\infty$, the system reaches a steady state:
\begin{align}
    &P_{\text{st}}(\eta_1=1,\eta_2,\dots,\eta_L)=\frac{\mu}{\binom{L-1}{N-1}\mu+\binom{L-1}{N}}\\
    &P_{\text{st}}(\eta_1=0,\eta_2,\dots,\eta_L)=\frac{1}{\binom{L-1}{N-1}\mu+\binom{L-1}{N}}.
\end{align}
Therefore, we obtain the site density and the two point correlation:
\begin{equation}
    \Braket{\eta_i}_N=\sum_{C}\eta_iP_{\text{st}}(C)=
    \begin{cases}
        \frac{\mu N}{\mu N+L-N} & (i=1)\\
        \\
        \frac{N\left[\mu(N-1)+L-N\right]}{(L-1)(\mu N+L-N)} & (i\neq 1)
    \end{cases},
    \label{site density}
\end{equation}
\begin{equation}
    \begin{split}
        \Braket{\eta_i\eta_j}_N&=\sum_{C}\eta_i\eta_jP_{\text{st}}(C)\\
        &=
        \begin{cases}
            \frac{\mu N(N-1)}{(L-1)(\mu N+L-N)} & (i=1,~j>i)\\
            \\
            \frac{N(N-1)\left[\mu (N-2)+L-N\right]}{(L-1)(L-2)(\mu N+L-N)} & (1<i<j\le L)
        \end{cases},
    \end{split}
    \label{two site correlation}
\end{equation}
where $\Braket{\cdot}_N$ means the ensemble average for the $N$-particles system.

\section{Derivation of the current fluctuation}\label{sec: result}
Here, we derive the second moment of the total integrated current $Q(t)$.
The total integrated current is defined as $Q(t)=\sum_{n=1}^N\delta x_n(t)$,
where $\delta x_n(t)$ is the displacement of the $n$th particle at time $t$.
The time evolution of the second moment of the total integrated current follows
\begin{equation}
    \begin{split}
        \frac{d\Braket{Q(t)^2}_N}{dt}=&\sum_{i=1}^L\frac{1}{\tau_i}\Braket{\eta_i(1-\eta_{i+1})}_N\\
        &+\left(1-\frac{1}{\mu}\right)\left[\Braket{Q(t)\eta_1\eta_2}_N-\Braket{Q(t)\eta_L\eta_1}_N\right],
    \end{split}
    \label{second moment of total integrated current derivation}
\end{equation}
which we derive in Appendix~\ref{sec: appendix current fluctuation}.
Therefore, the second moment of the total integrated current depends on the correlation between the total integrated current and the occupation number.
Hereafter, we consider $\kappa(N)\equiv\lim_{t\rightarrow\infty}d\Braket{Q(t)^2}_N/dt$.

\subsection{Correlation between total integrated current and occupation number}
Here, we derive the correlation between the total integrated current and the occupation number.
We note that the correlation between the total integrated current and the occupation number does not depend on the time,
which we prove in Appendix~\ref{sec: appendix correlation}.
In the long-time limit, we have $\Braket{Q\eta_i\eta_j}_N\equiv\lim_{t\rightarrow\infty}\Braket{Q(t)\eta_i\eta_j}_N$.

We consider $\Braket{Q\eta_i}_N$.
The time evolution of $\Braket{Q(t)\eta_i}_N$ for $3\le i \le L-1$ is
\begin{equation}
    \begin{split}
        &\lim_{t\rightarrow\infty}\frac{d\Braket{Q(t)\eta_i}_N}{dt}\\
        &=-\Braket{Q\eta_i}_N+\frac{1}{2}\Braket{Q\eta_{i+1}}_N+\frac{1}{2}\Braket{Q\eta_{i-1}}_N\\
        &=0.
    \end{split}
    \label{Qeta_i derivation}
\end{equation}
Thus, the recurrence relation holds
\begin{equation}
    \Braket{Q\eta_{i+1}}_N=2\Braket{Q\eta_i}_N-\Braket{Q\eta_{i-1}}_N.
\end{equation}
Using $\Braket{Q\eta_2}_N$ and $\Braket{Q\eta_3}_N$, we solve this recurrence relation and obtain
\begin{equation}
    \Braket{Q\eta_i}_N=i\left[\Braket{Q\eta_3}_N-\Braket{Q\eta_2}_N\right]
    +3\Braket{Q\eta_2}_N-2\Braket{Q\eta_3}_N.
    \label{Qeta_i}
\end{equation}

To obtain the correlation between the total integrated current and the occupation number, we derive $\Braket{Q\eta_2}_{L-1}$.
When $N=L-1$, there is a single hole.
The hole at site $i$ jump to the right site with rate $1/(2\tau_{i+1})$ and to the left site with rate $1/(2\tau_{i-1})$.
Using the result in Ref.~\cite{Derrida:1983aa},
the second moment of the total integrated current for $N=L-1$ is given by
\begin{align}
    \kappa(L-1)&=\frac{L^2}{\left(\sum_{i=1}^L\tau_i\tau_{i+1}\right)\left(\sum_{i=1}^L\tau_i^{-1}\right)}\label{QTM for L-1}\\
    &=\frac{L^2}{(2\mu+L-2)(\mu^{-1}+L-1)}.
\end{align}
Using Eq.~\eqref{second moment of total integrated current derivation}, we have
\begin{equation}
    \begin{split}
        &\Braket{Q\eta_1\eta_2}_{L-1}-\Braket{Q\eta_L\eta_1}_{L-1}\\
        &=-\frac{2(L-2)\mu(\mu-1)}{(2\mu+L-2)[\mu(L-1)+1]}.
    \end{split}
    \label{Qeta_1eta_2-Qeta_Leta_1}
\end{equation}
For $N=L-1$, it does not occur that two sites are empty.
Therefore, $\Braket{Q\eta_i\eta_j}_{L-1}$ can be represented as
\begin{equation}
    \Braket{Q\eta_i\eta_j}_{L-1}=\Braket{Q\eta_i}_{L-1}+\Braket{Q\eta_j}_{L-1}.
    \label{Qeta_ieta_j=Qeta_i+Qeta_j}
\end{equation}
Using this relation, Eq.~\eqref{Qeta_1eta_2-Qeta_Leta_1} is rewritten by
\begin{equation}
    \begin{split}
        &\Braket{Q\eta_2}_{L-1}-\Braket{Q\eta_L}_{L-1}\\
        &=-\frac{2(L-2)\mu(\mu-1)}{(2\mu+L-2)[\mu(L-1)+1]}.
    \end{split}
    \label{Qeta2 derivation}
\end{equation}
In order to derive the relation between $\Braket{Q\eta_2}_{L-1}$ and $\Braket{Q\eta_L}_{L-1}$, 
we consider the time evolution of $\Braket{Q(t)\eta_1}_N$:
\begin{equation}
    \begin{split}
        &\lim_{t\rightarrow\infty}\frac{d\Braket{Q(t)\eta_1}_N}{dt}\\
        =&\frac{1}{2}\Braket{Q\eta_2}_N+\frac{1}{2}\Braket{Q\eta_L}_N-\mu^{-1}\Braket{Q\eta_1}_N\\
        &-\frac{1-\mu^{-1}}{2}\left(\Braket{Q\eta_1\eta_2}_N+\Braket{Q\eta_L\eta_1}_N\right)\\
        =&0.
    \end{split}
    \label{Qeta_1 N}
\end{equation}
The average of the total integrated current is $0$ while site $1$ is occupied by a particle because sites except for site $1$ are homogeneous.
Thus, $\Braket{Q\eta_1}_N=0$.
For $N=L-1$, using Eq.~\eqref{Qeta_ieta_j=Qeta_i+Qeta_j}, Eq.~\eqref{Qeta_1 N} is represented as
\begin{equation}
    \Braket{Q\eta_2}_{L-1}+\Braket{Q\eta_L}_{L-1}=0.
\end{equation}
Using this relation and Eq.~\eqref{Qeta2 derivation}, $\Braket{Q\eta_2}_{L-1}$ is derived as
\begin{equation}
    \Braket{Q\eta_2}_{L-1}=-\frac{(L-2)\mu(\mu-1)}{(2\mu+L-2)[\mu(L-1)+1]}.
    \label{Qeta2}
\end{equation}

In order to derive $\Braket{Q\eta_i}_{L-1}$, we derive $\Braket{Q\eta_3}_{L-1}$.
The time evolution of $\Braket{Q(t)\eta_2}_N$ follows
\begin{equation}
    \begin{split}
        &\lim_{t\rightarrow\infty}\frac{d\Braket{Q(t)\eta_2}_N}{dt}\\
        =&\frac{1-\mu^{-1}}{2}\Braket{Q\eta_1\eta_2}_N+\frac{1}{2}\Braket{Q\eta_3}_N-\Braket{Q\eta_2}_N\\
        &+\frac{1}{2\mu}\Braket{\eta_1(1-\eta_2)}_N-\frac{1}{2}\Braket{\eta_3(1-\eta_2)}_N\\
        =&0.
    \end{split}
    \label{Qeta2 time differential}
\end{equation}
Using Eq.~\eqref{Qeta_ieta_j=Qeta_i+Qeta_j} and inserting Eq.~\eqref{Qeta2}
into Eq.~\eqref{Qeta2 time differential} for $N=L-1$, $\Braket{Q\eta_3}_{L-1}$ is derived as
\begin{equation}
    \Braket{Q\eta_3}_{L-1}=-\frac{(L-4)\mu(\mu-1)}{(2\mu+L-2)[\mu(L-1)+1]}.
    \label{Qeta3}
\end{equation}
Therefore, substituting Eqs.~\eqref{Qeta2} and \eqref{Qeta3} for Eq.~\eqref{Qeta_i}, $\Braket{Q\eta_i}_{L-1}$ for $2\le i\le L$ is represented by
\begin{equation}
    \Braket{Q\eta_i}_{L-1}=-\frac{(L+2-2i)\mu(\mu-1)}{(2\mu+L-2)[\mu(L-1)+1]}.
\end{equation}

We introduce two relations:
\begin{equation}
    f(N,L,\mu)\equiv\frac{\Braket{Q\eta_i}_N}{\Braket{Q\eta_i}_{L-1}}
    \label{f}
\end{equation}
and
\begin{equation}
    g(N,L,\mu)
    \equiv\frac{\Braket{Q\eta_1\eta_2}_N}{\Braket{Q\eta_2}_{N}}.
    \label{g}
\end{equation}
Equation~\eqref{f} indicates that the ratio of $\Braket{Q\eta_i}_N$ to $\Braket{Q\eta_i}_{L-1}$ for $i\neq 1,L/2+1$ does not depend on the site.
For site $1$ and $L/2+1$, $\Braket{Q\eta_1}_N=\Braket{Q\eta_{L/2+1}}_N=0$.
By considering the ratio of $\Braket{Q\eta_1\eta_2}_N$ to $\Braket{Q\eta_2}_N$, various properties emerge,
which will be discussed later.
Substituting Eq.~\eqref{f} for Eq.~\eqref{Qeta_1 N}, we have
\begin{equation}
    \Braket{Q\eta_1\eta_2}_N+\Braket{Q\eta_L\eta_1}_N=0.
\end{equation}
Inserting Eqs.~\eqref{f} and \eqref{g} into Eq.~\eqref{Qeta2 time differential}, the relation between $f(N,L,\mu)$ and $g(N,L,\mu)$ becomes
\begin{equation}
    \begin{split}
        &f(N,L,\mu)\\
        &=\frac{\Braket{\eta_3(1-\eta_2)}_N-\mu^{-1}\Braket{\eta_1(1-\eta_2)}_N}{\Braket{Q\eta_2}_{L-1}\left[(1-\mu^{-1})g(N,L,\mu)-\frac{L}{L-2}\right]}.
    \end{split}
\end{equation}
Therefore, $\Braket{Q\eta_1\eta_2}_N$ is represented as
\begin{equation}
    \begin{split}
        &\Braket{Q\eta_1\eta_2}_N\\
        &=\frac{\Braket{\eta_3(1-\eta_2)}_N-\mu^{-1}\Braket{\eta_1(1-\eta_2)}_N}{(1-\mu^{-1})g(N,L,\mu)-\frac{L}{L-2}}g(N,L,\mu).
    \end{split}
\end{equation}

We consider the property of $g(N,L,\mu)$.
$\Braket{Q\eta_1\eta_2}_N$ is represented as
\begin{equation}
    \Braket{Q\eta_1\eta_2}_N=\Braket{Q\eta_2}_N+\Braket{Q(1-\eta_1)(1-\eta_2)}_N.
\end{equation}
This relation divided by $\Braket{Q\eta_2}_N$ yields
\begin{equation}
    g(N,L,\mu)=1+\frac{\Braket{Q(1-\eta_1)(1-\eta_2)}_N}{\Braket{Q\eta_2}_N}.
\end{equation}
For $N=L-1$, it does not occur that two sites are empty.
Namely, $\Braket{Q(1-\eta_1)(1-\eta_2)}_{L-1}=0$, i.e., we obtain $g(L-1,L,\mu)=1$.
For $\mu\rightarrow\infty$, a particle constantly occupies site $1$ and becomes the reflective wall.
Thus, $\lim_{\mu\rightarrow\infty}\Braket{Q(1-\eta_1)(1-\eta_2)}_N=0$, that is, $\lim_{\mu\rightarrow\infty}g(N,L,\mu)=1$.
Therefore, the function $g(N,L,\mu)$ satisfying $g(L-1,L,\mu)=1$ and $\lim_{\mu\rightarrow\infty}g(N,L,\mu)=1$ is given by
\begin{equation}
    g(N,L,\mu)
    \cong\frac{a(L,\mu)(L-1-N)+\mu}{b(L,\mu)(L-1-N)+\mu},
    \label{g approximate}
\end{equation}
where $a(L,\mu)=o(\mu)$ and $b(L,\mu)=o(\mu)$.
We assume that $a(L,\mu)$ and $b(L,\mu)$ do not depend on $N$.
We represent the parameters $a(L,\mu)$ and $b(L,\mu)$ by $g(2,L,\mu)$ and $g(L-2,L,\mu)$,
\begin{align}
    &a(L,s)
    =g(L-2,L,\mu)(b(L,\mu)+\mu)-\mu\\
    &b(L,\mu)
    =\frac{\mu(L-4)\left[1-g(2,L,\mu)\right]}{(L-3)\left[g(L-2,L,\mu)-g(2,L,\mu)\right]}-\mu.
\end{align}

\subsection{Second moment of total integrated current}
\begin{figure}[htbp]
    \centering
    \subfigure[][$\mu=10$]{\includegraphics[width=8.6cm]{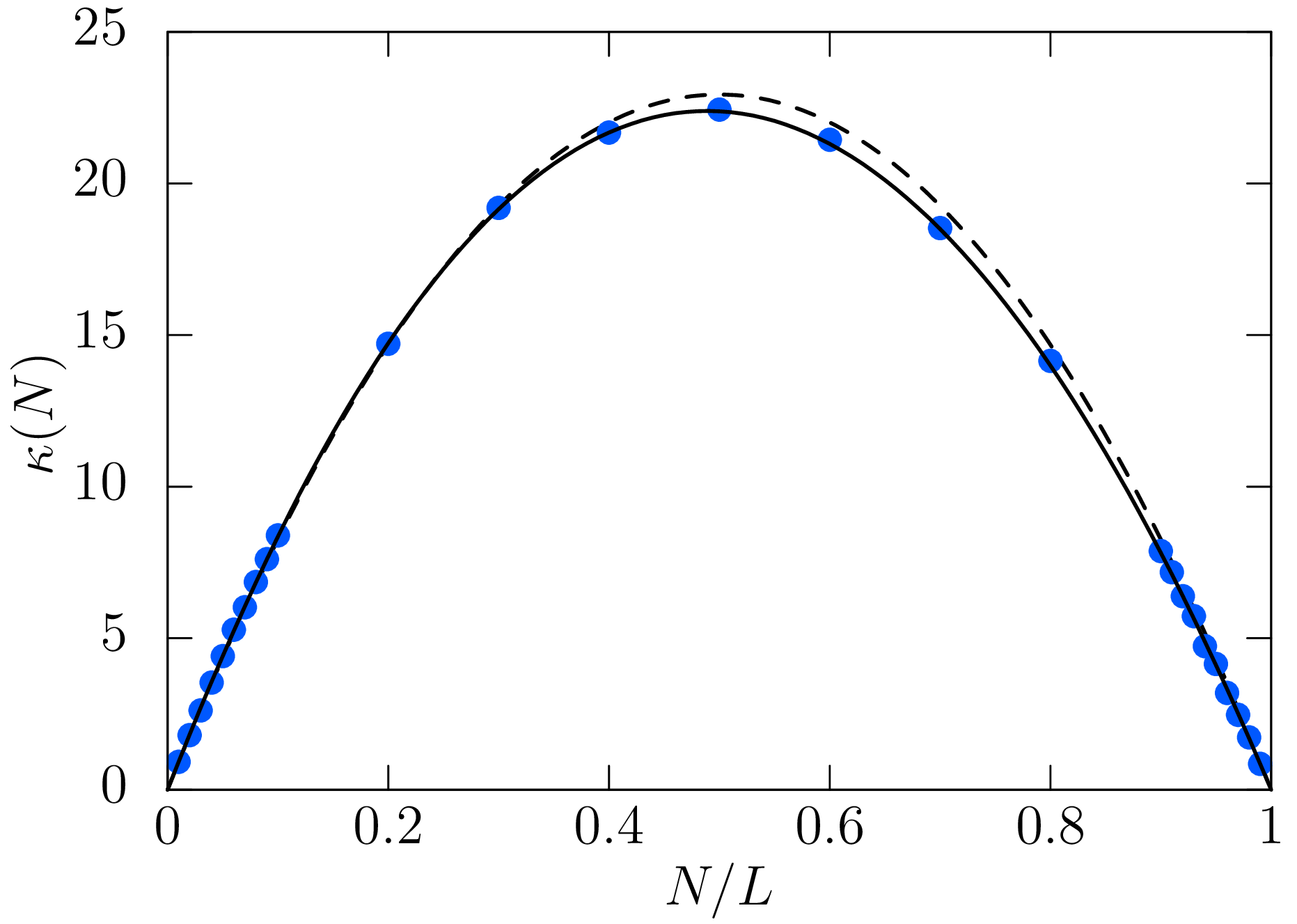}}
    \subfigure[][$\mu=100$]{\includegraphics[width=8.6cm]{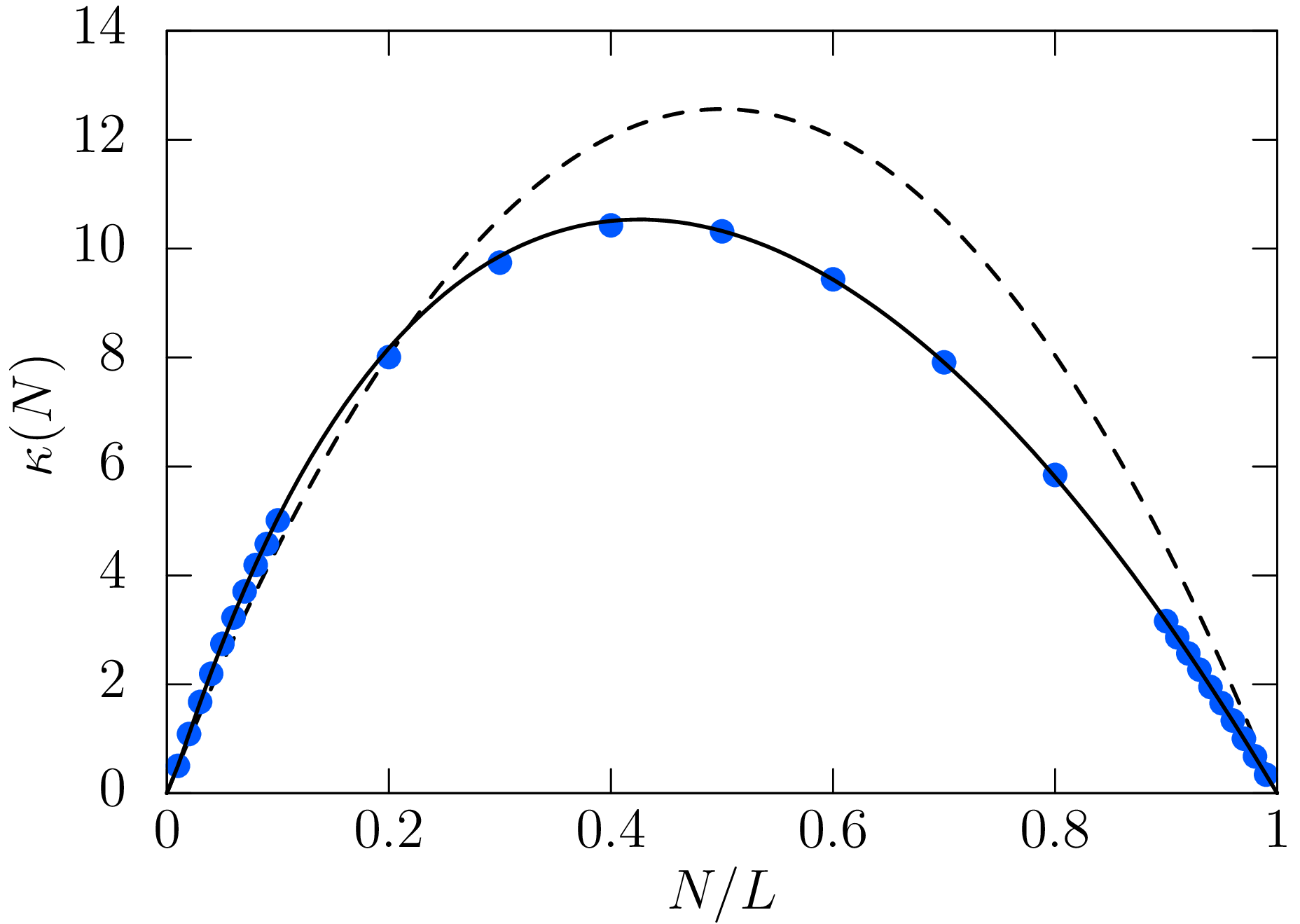}}
    \caption{Relation between $\kappa(N)$ and density for the SSEP with a defect site ($L=100$).
    Symbols are the results of the numerical simulation of the dynamics of the SSEP with a defect site.
    Solid lines represent Eq.~\eqref{Second moment of total integrated current for single defect system}, where we use $g(2,L,\mu)$ and $g(L-2,L,\mu)$ obtained by the numerical simulations.
    Dashed lines show the result in the homogeneous system, Eq.~\eqref{current fluctuations for homogeneous system},
    where $\tau$ is set to equal to the average of the waiting time of the SSEP with a defect site, i.e., $\tau=(L-1+\mu)/L$.}
    \label{fig: single_defect_system}
\end{figure}

Based on the results of the previous section, the second moment of the total integrated current is represented by
\begin{equation}
    \begin{split}
        \kappa(N)=&\frac{N(L-N)}{(L-1)(\mu N+L-N)}\biggl[(\mu-1)(N-1)\\
        &+L-\frac{2(\mu-1)^2(N-1)g(N,L,\mu)}{\mu L-(\mu-1)(L-2)g(N,L,\mu)}\biggr].
    \end{split}
    \label{Second moment of total integrated current for single defect system}
\end{equation}
We note that the function $g(N,L,\mu)$ follows Eq.~\eqref{g approximate}, and $g(2,L,\mu)$ and $g(L-2,L,\mu)$ are obtained by the numerical simulations.
We compare the theoretical result, Eq.~\eqref{Second moment of total integrated current for single defect system},
with the numerical result.
Figure~\ref{fig: single_defect_system} shows good agreement between numerical simulations and the theory.
The density dependence of $\kappa(N)$ for the SSEP with a defect site is different from that for the homogeneous system \eqref{current fluctuations for homogeneous system}.
In particular, it becomes asymmetric around $N/L=1/2$.
Furthermore, in the intermediate and high-density regime, $\kappa(N)$ for the SSEP with a defect site is smaller than that for the homogeneous system.
This result is that the defect hiders the particle movement, leading to a reduction in fluctuations.
On the other hand, in the low-density regime, $\kappa(N)$ for the SSEP with a defect site exceeds that for the homogeneous system.
This property is counterintuitive because despite the defect impeding the movement of particles, the fluctuations increase.
In terms of diffusivity, adjusting the defect allows for control over the particle's diffusivity.

We examine the asymmetry in the density dependence of $\kappa(N)$.
In the homogeneous system, the symmetry of the density dependence of $\kappa(N)$
arises from the identical dynamics of particles and holes.
In the system with a defect site, the dynamics of holes differ from those of particles at site $i=1,2,L$, while remaining the same elsewhere.
As a result, the disparate dynamics of particles and holes induce asymmetry.

\section{Application: SSEP on a quenched random energy landscape}\label{sec: Application}
\begin{figure*}
    \centering
    \subfigure[\hspace{6.2cm}][Exponential distribution]{\includegraphics[width=4.3cm]{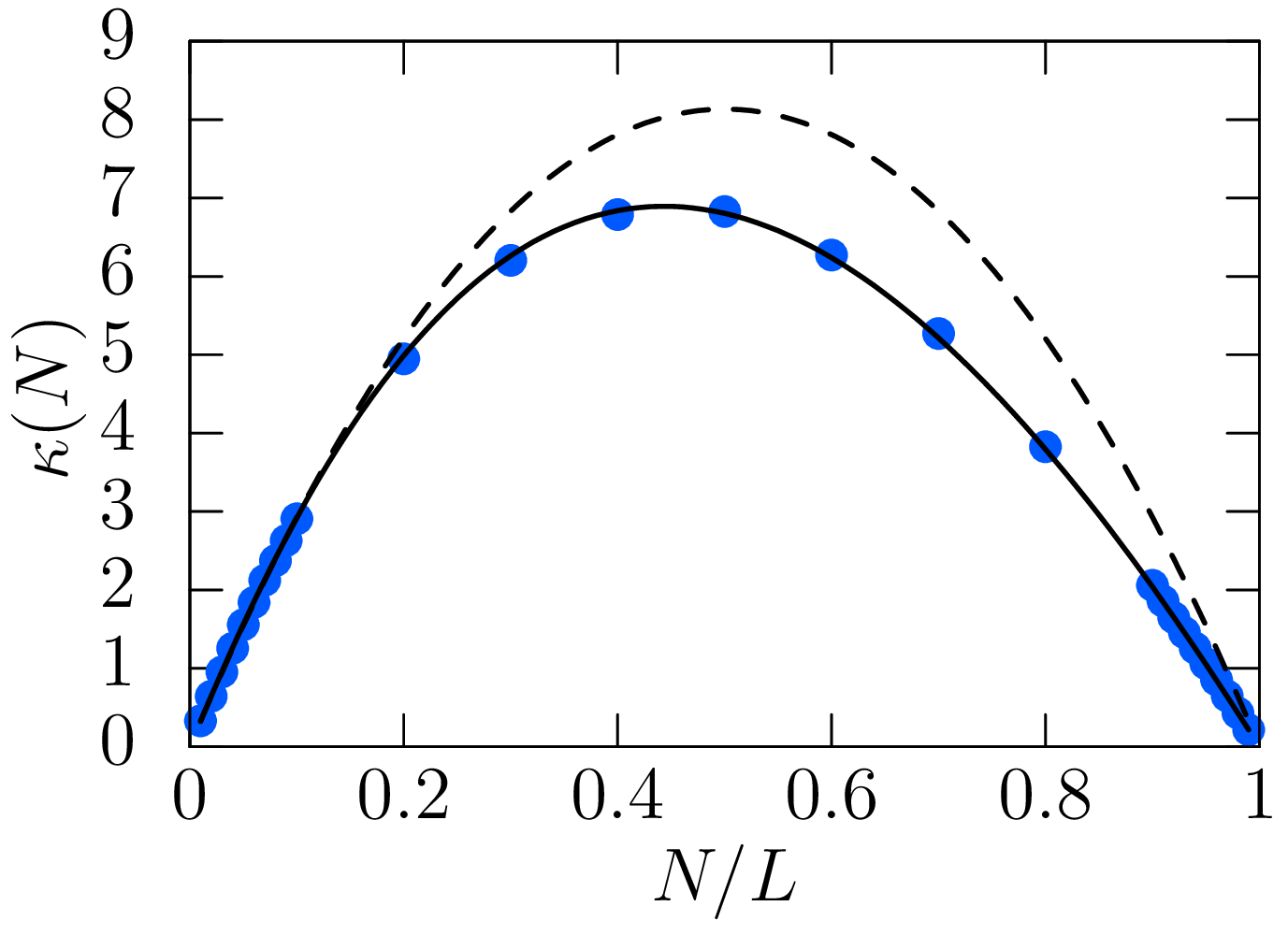}}
    \subfigure[\hspace{6.2cm}][Gamma distribution]{\includegraphics[width=4.3cm]{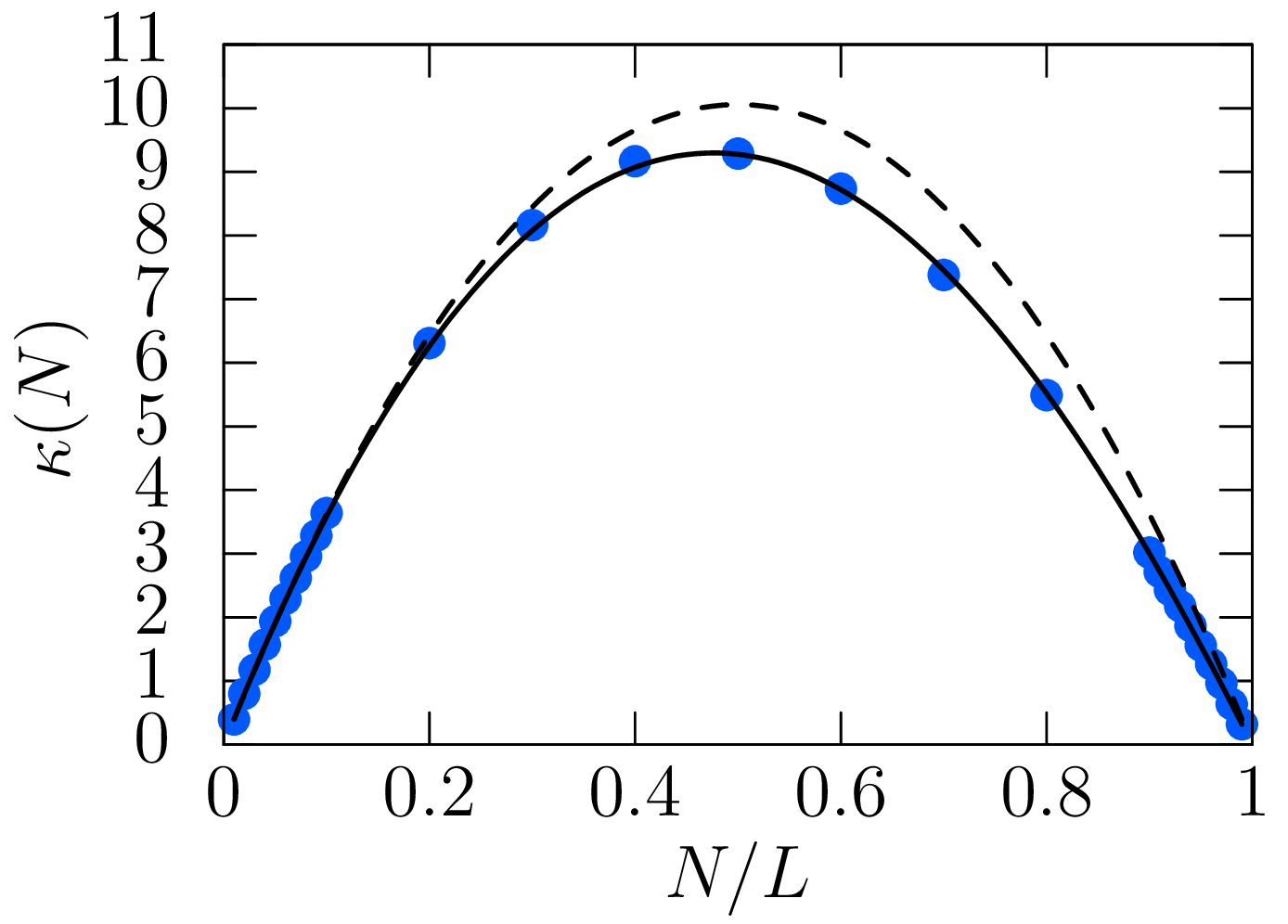}}
    \subfigure[\hspace{6.2cm}][Half-gaussian distribution]{\includegraphics[width=4.3cm]{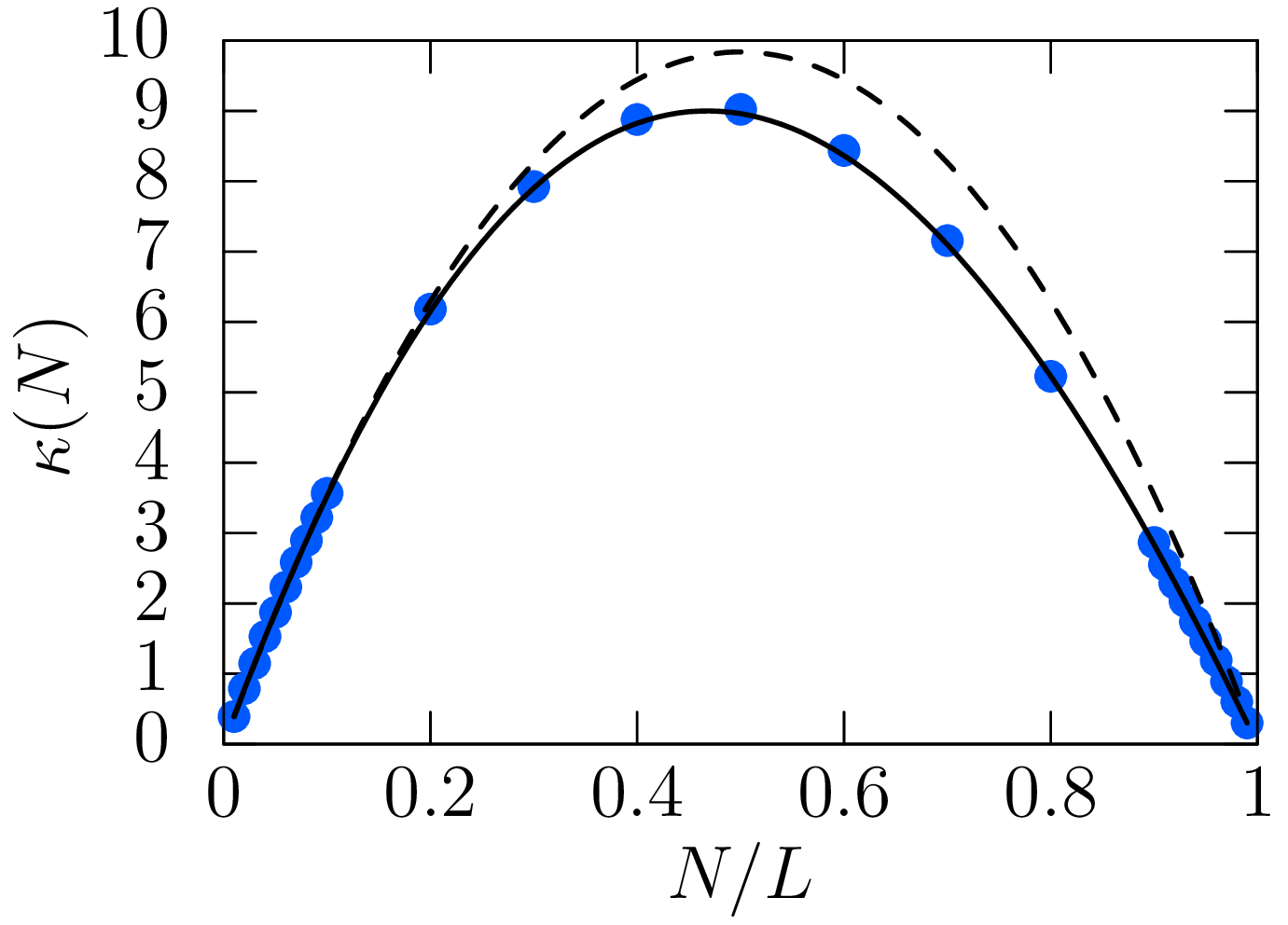}}
    \subfigure[\hspace{6.2cm}][Uniform distribution]{\includegraphics[width=4.3cm]{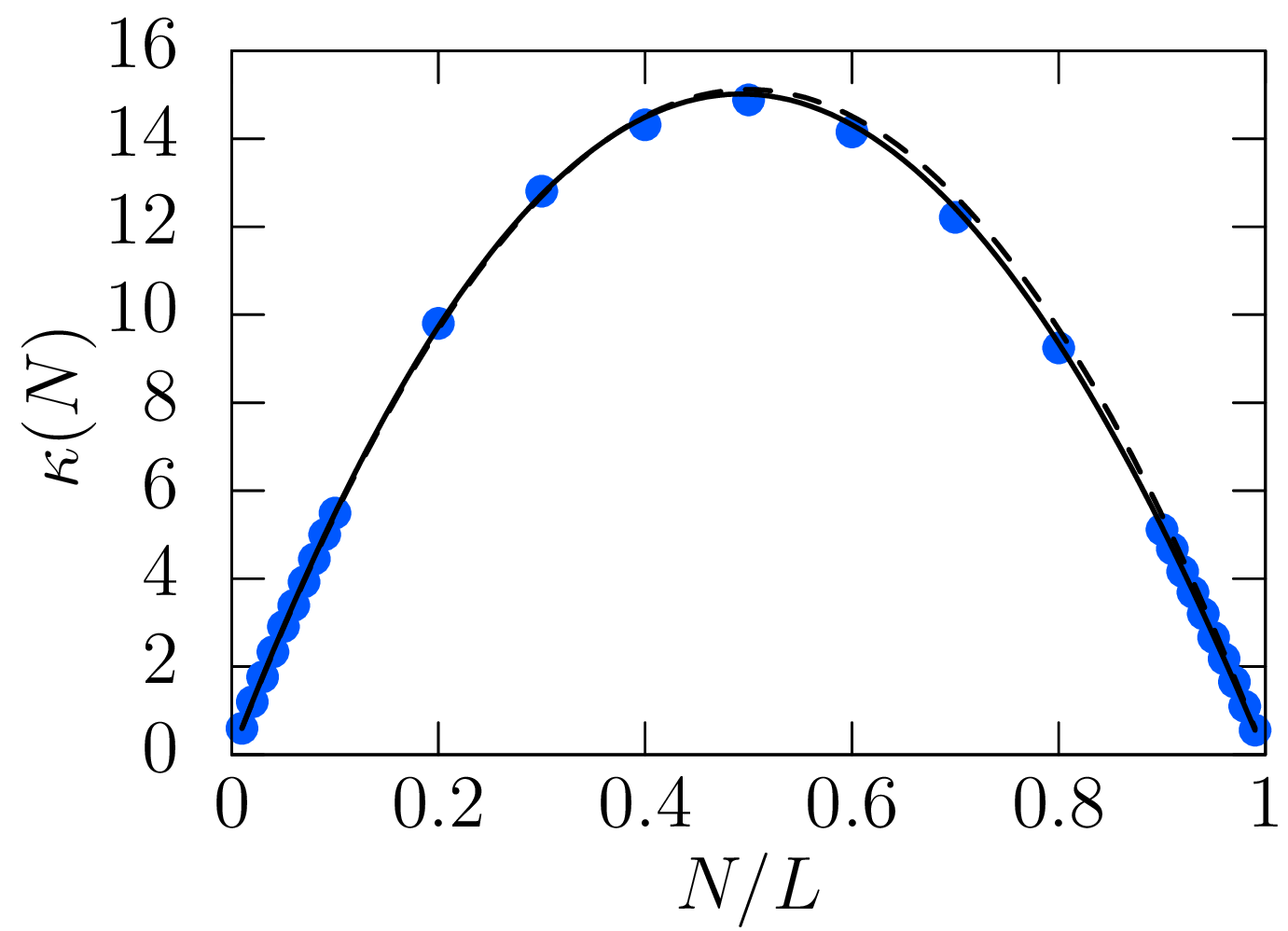}}\\
    \subfigure[\hspace{6.2cm}][Exponential distribution]{\includegraphics[width=4.3cm]{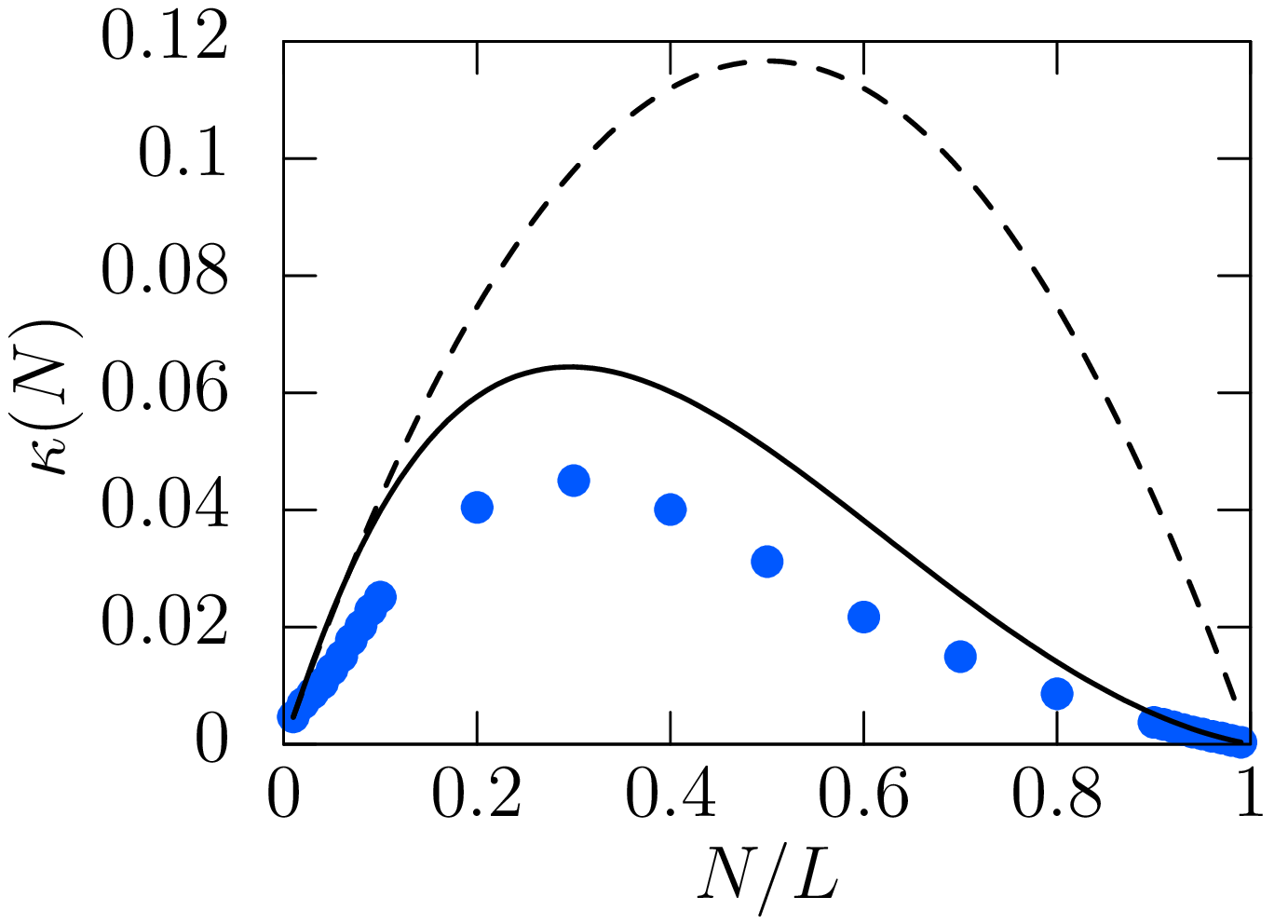}}
    \subfigure[\hspace{6.2cm}][Gamma distribution]{\includegraphics[width=4.3cm]{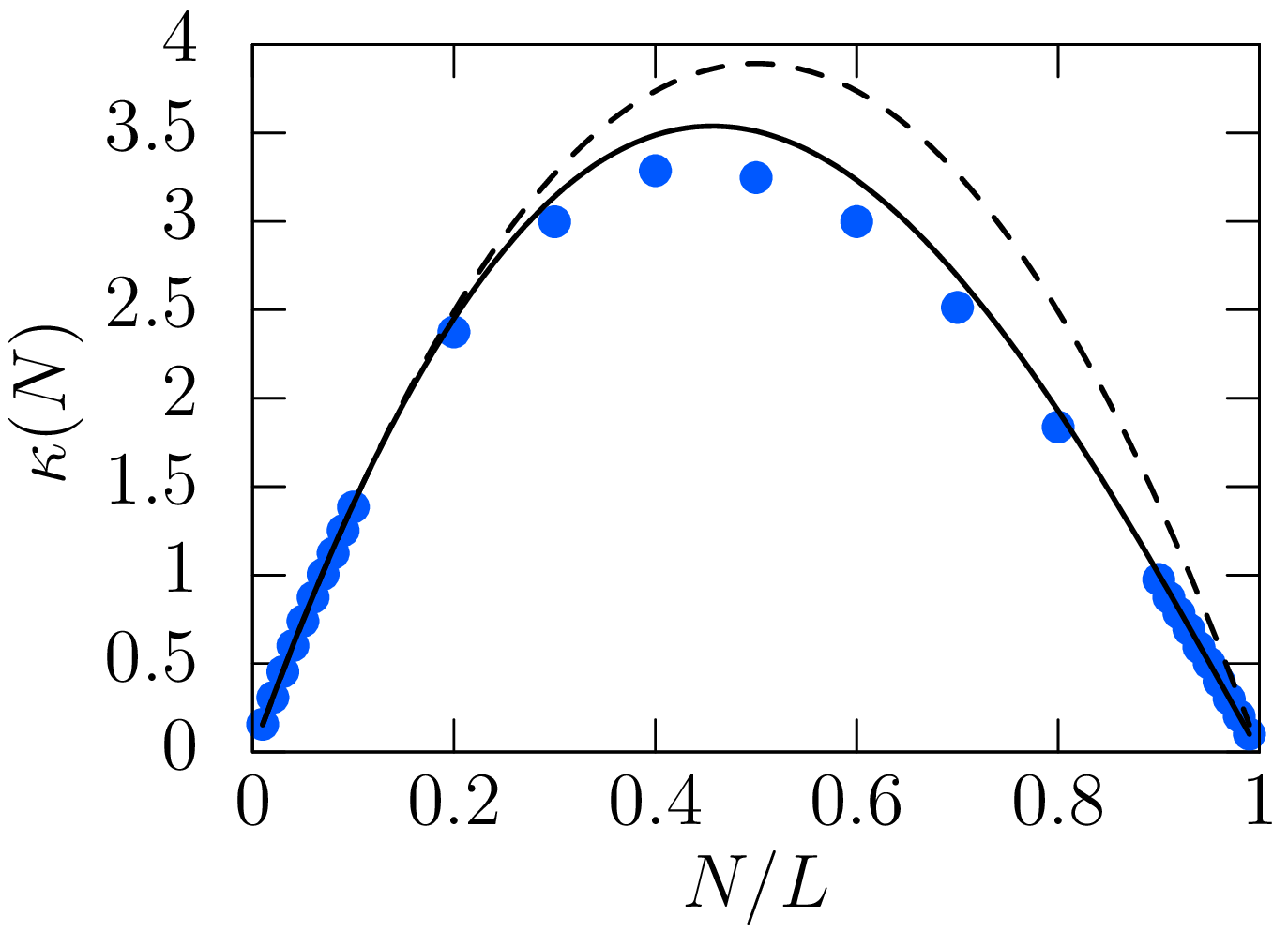}}
    \subfigure[\hspace{6.2cm}][Half-gaussian distribution]{\includegraphics[width=4.3cm]{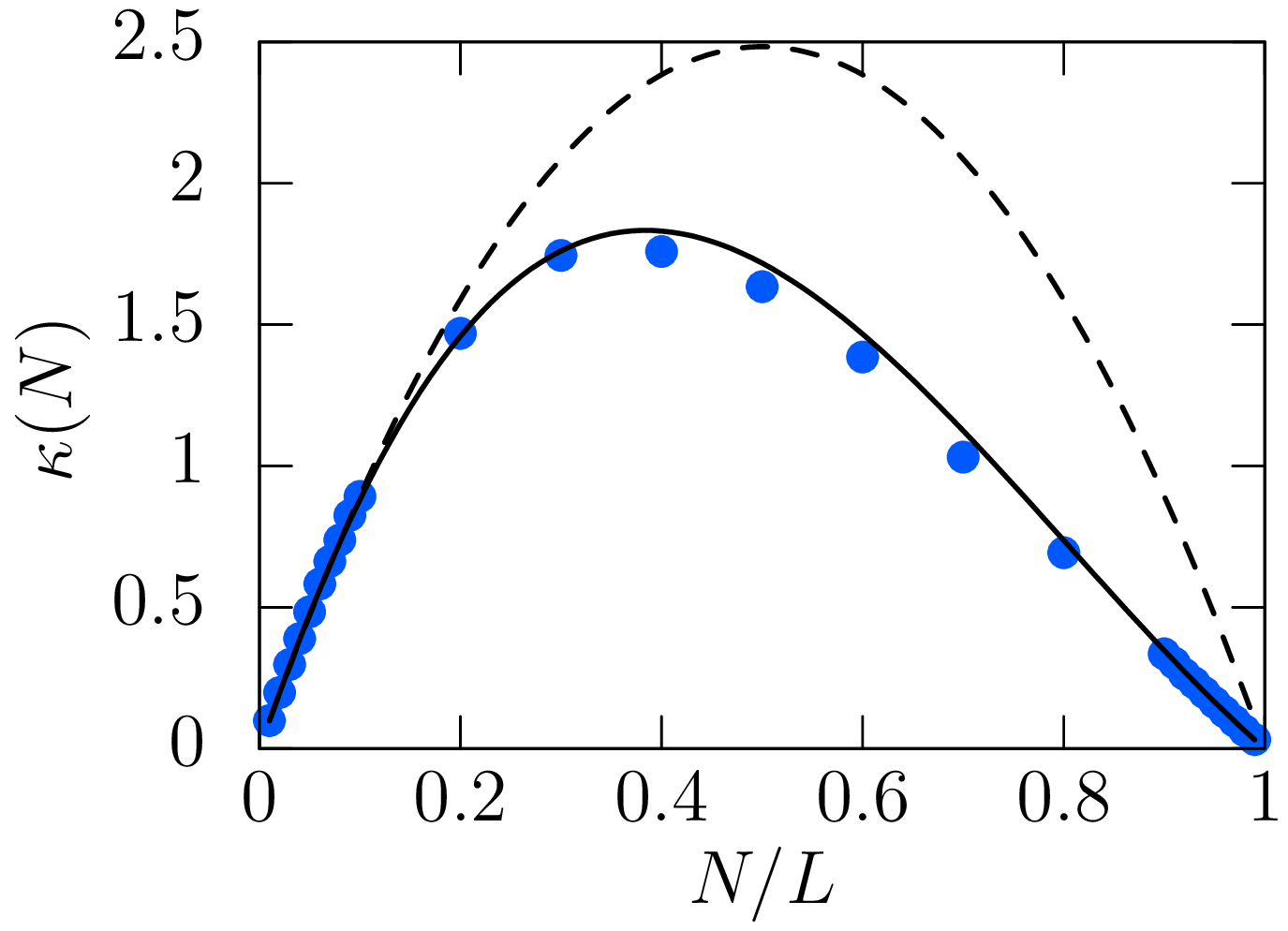}}
    \subfigure[\hspace{6.2cm}][Uniform distribution]{\includegraphics[width=4.3cm]{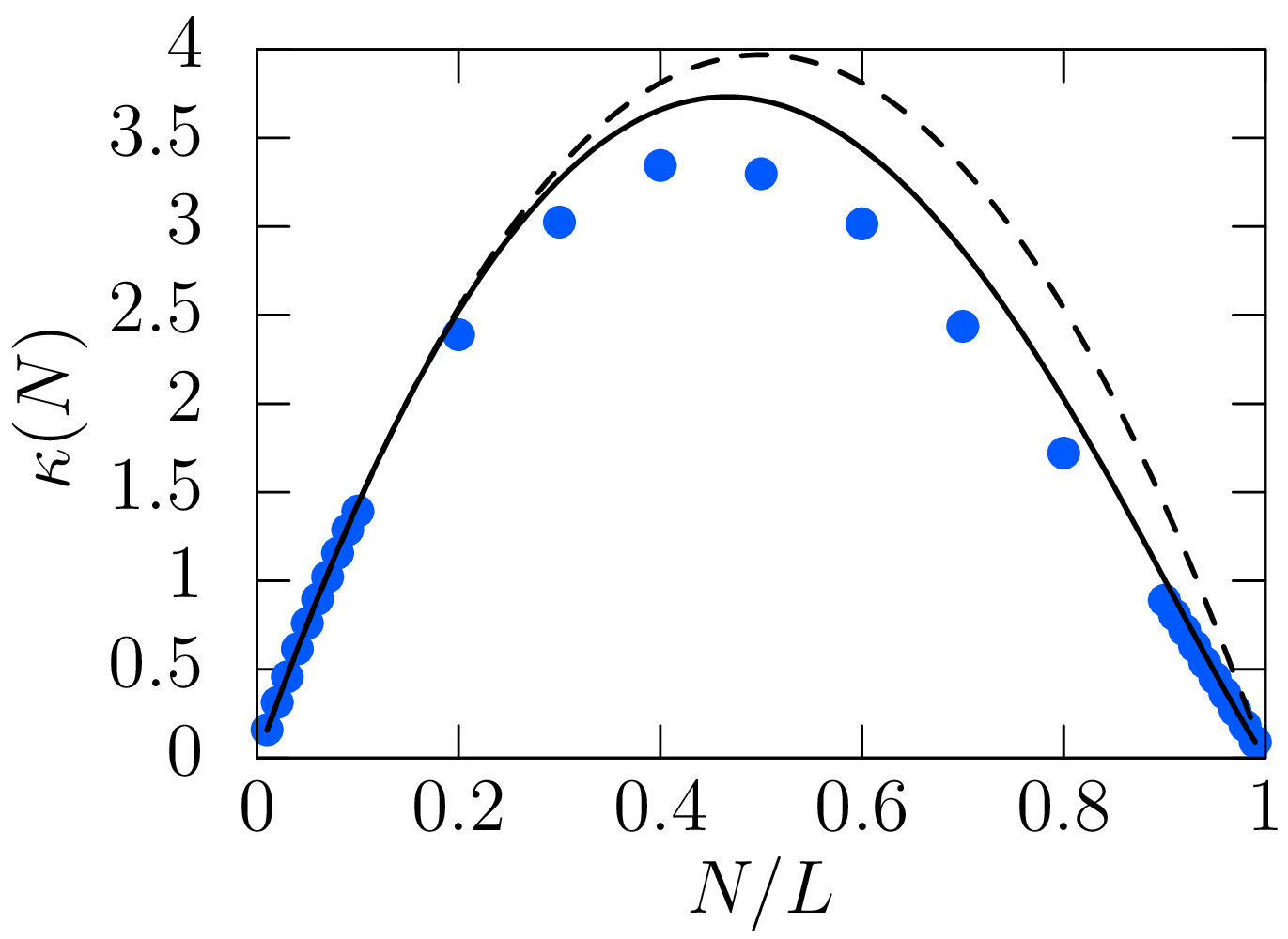}}
    \caption{Relation between $\kappa(N)$ and density, corresponding to the SSEP on a quenched random energy landscape ($L=100$, $T=1$, $\tau_c=1$):
    (a)-(d) $\Braket{E}<T$ and (e)-(h) $\Braket{E}>T$.
    (a),(e) The energy depth distribution is the exponential distribution ($T/T_g=1.5$ for (a) and $T/T_g=0.5$ for (e)).
    (b),(f) The energy depth distribution is the gamma distribution ($T/T_g=2.5,~k=2$ for (b) and $T/T_g=2.5,~k=4$ for (f)).
    (c),(g) The energy depth distribution is the half-gaussian distribution ($\sigma^2=1$ for (c) and $\sigma^2=4$ for (g)).
    (d),(h) The energy depth distribution is the uniform distribution ($\varepsilon=1$ for (d) and $\varepsilon=3$ for (h)).
    Circles are the numerical simulation of the dynamics of the SSEP on a quenched random energy landscape.
    Solid lines represent Eq.~\eqref{Second moment of total integrated current for QTM}, where we use $g(2,L,\mu)$ and $g(L-2,L,\mu)$ obtained by the numerical simulations of the dynamics of the SSEP with a defect site.
    Dashed lines show the result in the homogeneous system, Eq.~\eqref{current fluctuations for homogeneous system},
    where $\tau$ is set to equal to the average of the waiting time of the SSEP on the quenched random energy landscape, i.e., $\tau=\sum_{i=1}^L\tau_i/L$.
    }
    \label{fig: Second moment of total integrated current for QTM}
\end{figure*}

Here, we apply our theory of the current fluctuations of the SSEP with a defect site to the SSEP on a quenched random energy landscape.
Quenched disorder means that when realizing the random energy landscape, it does not change with time.
The quenched random energy landscape for the SSEP assigns random energy traps to sites, influencing particle jump rates.
More precisely, at each site, the depth $E>0$ of the energy trap is randomly assigned.
The energies of the traps are IID random variables following distribution $\phi(E)$.
A particle escapes from the trap and attempts to jump to the nearest neighbors.
Waiting times at site $i$ are IID random variables with an exponential distribution, $\psi^{(i)}(\tau)=\tau_i^{-1}\exp(-\tau/\tau_i)$.
The mean waiting time follows the Arrhenius law, $\tau_i=\tau_c\exp(E_i/T)$,
where $E_i$ is the depth of the energy at site $i$, $T$ the temperature, and $\tau_c$ typical time.

We consider four distributions of the energy depth $\phi(E)$:
the exponential distribution
\begin{equation}
    \phi(E)=\frac{1}{T_g}\exp\left(-\frac{E}{T_g}\right)\quad (E\geq0),
    \label{exponential distribution}
\end{equation}
where $T_g$ is the glass temperature,
the gamma distribution
\begin{equation}
    \phi(E)=\frac{T_g^{-k}E^{k-1}}{\Gamma(k)}\exp\left(-\frac{E}{T_g}\right)\quad (E\geq0),
\end{equation}
where $k$ is a scale parameter,
the half-gaussian distribution
\begin{equation}
    \phi(E)=\frac{\sqrt{2}}{\sigma\sqrt{\pi}}\exp\left(-\frac{E^2}{2\sigma^2}\right)\quad (E\geq0),
\end{equation}
where $\sigma$ is a scale parameter and the mean of this distribution is $\sigma\sqrt{2/\pi}$,
and the uniform distribution
\begin{equation}
    \phi(E)=
    \begin{cases}
        \varepsilon^{-1} & (0\le E <\varepsilon)\\
        0 & (\mathrm{otherwise})
    \end{cases}.
\end{equation}
Based on the Arrhenius law, the probability that the mean waiting time $\tau$ is smaller than $x$ is given by $\Pr(\tau\le x)\cong \Pr(E\le T\ln(x/\tau_c))$.
Hence, the distributions of the mean waiting times follow
\begin{equation}
    \psi(\tau)\cong\alpha\tau_c^{\alpha}\tau^{-(1+\alpha)}
    \quad (\tau\geq\tau_c)
    \label{power law distribution}
\end{equation}
when $\phi(E)$ is the exponential distribution,
where $\alpha\equiv T/T_g$,
\begin{equation}
    \psi(\tau)\cong\frac{\alpha^k\tau_c^{\alpha}}{\Gamma(k)}\left[\ln\left(\frac{\tau}{\tau_c}\right)\right]^{k-1}\tau^{-(1+\alpha)}
    \quad (\tau\geq\tau_c)
\end{equation}
when $\phi(E)$ is the gamma distribution,
\begin{equation}
    \psi(\tau)\cong\sqrt{\frac{2}{\pi}}\frac{T}{\sigma\tau}\exp\left[-\left(\frac{T\ln (\tau/\tau_c)}{\sigma}\right)^2\right]
    \quad (\tau\geq\tau_c)
\end{equation}
when $\phi(E)$ is the half-gaussian distribution,
and
\begin{equation}
    \psi(\tau)\cong\frac{T}{\varepsilon}\tau^{-1}
    \quad \left(\tau_c\le \tau<\tau_ce^{\varepsilon/T}\right)
\end{equation}
when $\phi(E)$ is the uniform distribution.

We derive the approximate value of $\kappa(N)$ for the SSEP on a quenched random energy landscape.
We consider that the longest mean waiting time contributes significantly $\kappa(N)$.
We employ a partial-mean-field approach, where the SSEP on a quenched random energy landscape is mapped to the SSEP with a defect site.
In the SSEP with a defect site, the mean waiting times at sites are defined as
\begin{equation}
    \tilde{\tau}_i=\begin{cases}
        \tau_{\max}=\max\{\tau_1,\dots,\tau_L\} & (i=1)\\
        \nu=\frac{1}{L-1}\left[\sum_{j=1}^L\tau_j-\tau_{\max}\right] & (i\neq 1)
    \end{cases}.
\end{equation}
Hence, the mean waiting time at site $1$ is the longest one of the SSEP on a quenched random energy landscape,
and the others are the sample average of the mean waiting times except for the longest one.
Based on $\kappa(N)$ of the SSEP with a defect site \eqref{Second moment of total integrated current for single defect system},
we obtain $\kappa(N)$ of the SSEP on a quenched random energy landscape
\begin{equation}
    \begin{split}
        \kappa(N)\cong&\frac{N(L-N)}{\nu(L-1)(\mu N+L-N)}\biggl[(\mu-1)(N-1)\\
        &+L-\frac{2(\mu-1)^2(N-1)\chi g(N,L,\mu)}{\mu L-(\mu-1)(L-2)\chi g(N,L,\mu)}\biggr],
    \end{split}
    \label{Second moment of total integrated current for QTM}
\end{equation}
where $\mu=\tau_{\max}/\nu$ and $\chi$ is the correction term.
$\kappa(L-1)$ has to obey Eq.~\eqref{QTM for L-1}.
Therefore, $\chi$ is represented as
\begin{equation}
    \chi=
    \frac{\mu L}{(\mu-1)(L-2)}\frac{\frac{\mu(L-2)+2}{\mu(L-1)+1}-\frac{\nu L^2}{\left(\sum_{i=1}^L\tau_i\tau_{i+1}\right)\left(\sum_{i=1}^L\tau_i^{-1}\right)}}{\frac{\mu L}{\mu(L-1)+1}-\frac{\nu L^2}{\left(\sum_{i=1}^L\tau_i\tau_{i+1}\right)\left(\sum_{i=1}^L\tau_i^{-1}\right)}}.
\end{equation}

Figure~\ref{fig: Second moment of total integrated current for QTM} shows the comparison between the numerical result
and the approximate result \eqref{Second moment of total integrated current for QTM}.
We classify the two cases:
$\Braket{E}<T$ (Figs.~\ref{fig: Second moment of total integrated current for QTM}(a)-(d)) and $\Braket{E}>T$ (Figs.~\ref{fig: Second moment of total integrated current for QTM}(e)-(h)),
where $\Braket{E}\equiv\int_{0}^{\infty}E\phi(E)dE$.
The dependence of $\kappa(N)$ on the density becomes asymmetric concerning $N/L=1/2$
because the hole dynamics differ from the particle ones.
For $\Braket{E}<T$, the approximate results coincide with the numerical results.
Therefore, $\kappa(N)$ crucially depends on the longest waiting time, and the waiting times except for the longest one can be homogenized using their sample average.
Conversely, for $\Braket{E}>T$, we can estimate the density at which $\kappa(N)$ is maximum [Fig.~\ref{fig: Second moment of total integrated current for QTM}(e)],
and the approximate results correspond to the numerical results in the low- and high-density limit [Figs.~\ref{fig: Second moment of total integrated current for QTM}(f)-(h)].
However, the approximate results deviate from the numerical results in the intermediate density range.
Consequently, the homogenization of the waiting times excluding the longest one is not successful in the intermediate density range.

We compare the numerical results with the theoretical result \eqref{current fluctuations for homogeneous system} in the homogeneous environment [see Fig.~\ref{fig: Second moment of total integrated current for QTM}].
In the intermediate and high-density regime, $\kappa(N)$ for the SSEP on the quenched random energy landscape is consistently smaller than that for the homogeneous environment.
In the low-density regime, $\kappa(N)$ for the SSEP on the quenched random energy landscape is approximately equal to that for the homogeneous environment [Figs.~\ref{fig: Second moment of total integrated current for QTM}(a)-(d), (f)-(h)]
or smaller than it [Fig.~\ref{fig: Second moment of total integrated current for QTM}(e)].
$\kappa(N)$ for the SSEP on the quenched random energy landscape does not exceed that in the homogeneous environment
because the heterogeneity in the quenched random energy landscape is stronger than that in the system with a defect site.

\section{Discussion}\label{sec: discussion}
\begin{figure}[htbp]
    \centering
    \includegraphics[width=8.6cm]{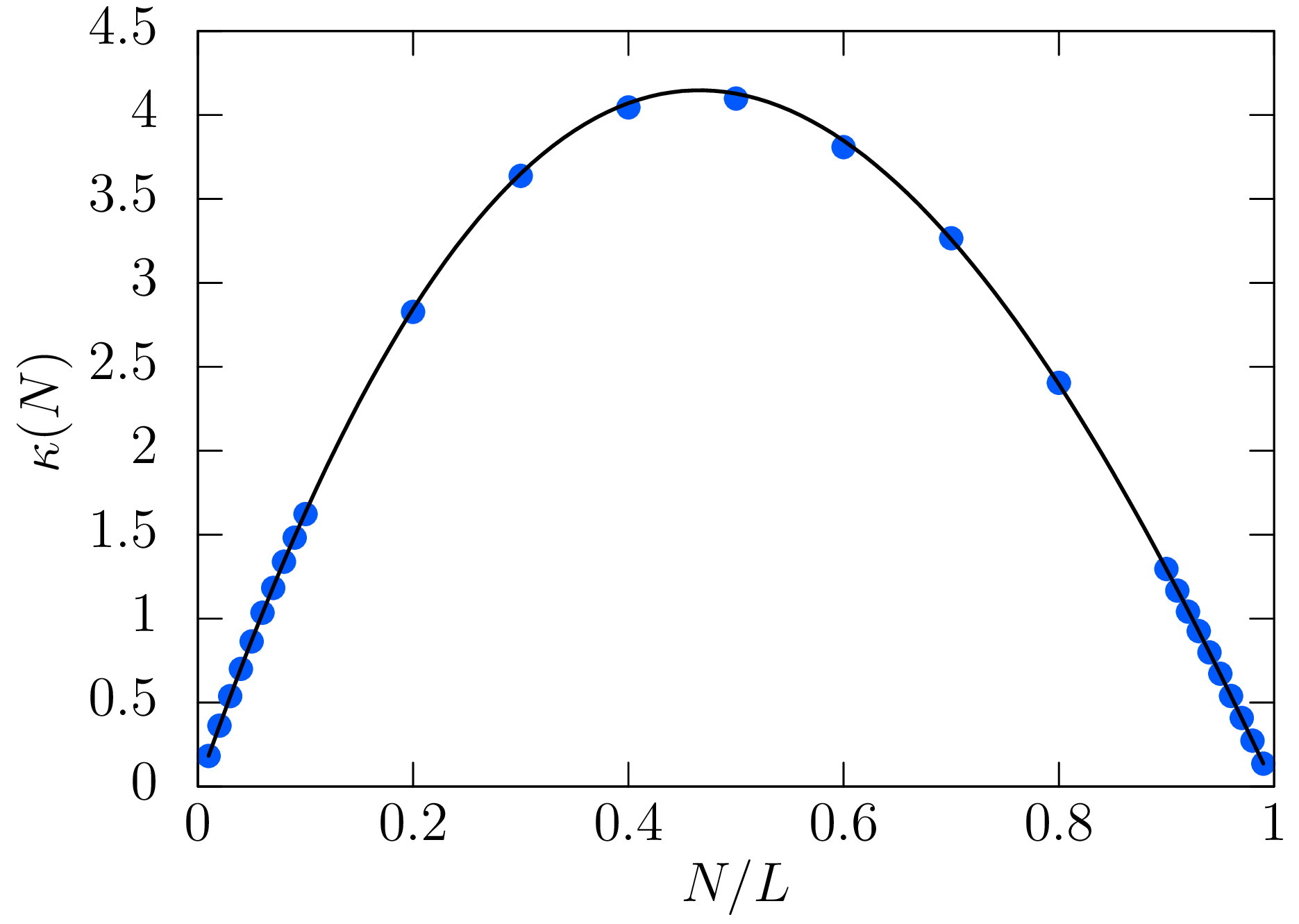}
    \caption{Relation between $\kappa(N)$ and density, corresponding to the SSEP on a quenched random energy landscape ($L=100$, $T=1$, $\tau_c=1$).
    Replacing the second to eighth longest waiting times in the system of Fig.~\ref{fig: Second moment of total integrated current for QTM}(f) with the mean waiting time $\Braket{\tau}$.
    Circles are the numerical simulation of the dynamics of the SSEP on a quenched random energy landscape.
    Solid lines represent Eq.~\eqref{Second moment of total integrated current for QTM}, where we use $g(2,L,\mu)$ and $g(L-2,L,\mu)$ obtained by the numerical simulations of the dynamics of the SSEP with a defect site.}
    \label{fig: Second moment of total integrated current for QTM remove method}
\end{figure}

We discuss that for $\Braket{E}>T$, the approximate result \eqref{Second moment of total integrated current for QTM} does not agree with the numerical result in the intermediate density range [see Figs.~\ref{fig: Second moment of total integrated current for QTM}(e)-(h)].
We presume that the disagreement between the numerical result and the approximate result is induced by ignoring the contributions of longer waiting times, such as the second longest waiting time.
Figure~\ref{fig: Second moment of total integrated current for QTM remove method} shows the result
when longer waiting times in the system are replaced with the mean waiting time $\Braket{\tau}\equiv\int_{0}^{\infty}\tau\psi(\tau)d\tau$.
The approximate result \eqref{Second moment of total integrated current for QTM} coincides with the numerical result.
As a result, for $\Braket{E}>T$, the current fluctuation crucially depends not only on the longest waiting time but also on other longer waiting times, such as the second longest waiting time.
Therefore, the dependence of the current fluctuation on the energy landscape for $\Braket{E}>T$ is even stronger than that for $\Braket{E}<T$.

We compare the current fluctuation in the single-particle system and the many-particle system.
In the single-particle system, the current fluctuation is the same as the mean square displacement, i.e., $\kappa(1)=\bar{\tau}^{-1}$,
where $\bar{\tau}=\sum_{i=1}^L\tau_i/L$ \cite{Derrida:1983aa,AkimotoBarkaiSaito,AkimotoBarkaiSaito2018}.
When the mean waiting time $\Braket{\tau}$ is finite, we have $\bar{\tau}\rightarrow\Braket{\tau}$ ($L\rightarrow\infty$) by the law of large numbers.
Hence, in the large-$L$ limit, $\kappa(1)$ does not depend on the energy landscape.
When the mean waiting time diverges (e.g., the power law distribution \eqref{power law distribution} for $\alpha<1$),
the law of large numbers does not hold, and $\kappa(1)$ shows the sample-to-sample fluctuation \cite{AkimotoBarkaiSaito,AkimotoBarkaiSaito2018}.
In the many-particle system, the current fluctuation $\kappa(N)$ depends on the energy landscape.
Thus, $\kappa(N)$ in the intermediate density range exhibits the sample-to-sample fluctuation.
As a result of the many-body interactions, the sample-to-sample fluctuation of $\kappa(N)$ emerges in the intermediate density range,
even in a one-particle system that does not exhibit the sample-to-sample fluctuation.

\begin{figure}
    \centering
    \subfigure[\hspace{4cm}]{\includegraphics[width=4.2cm]{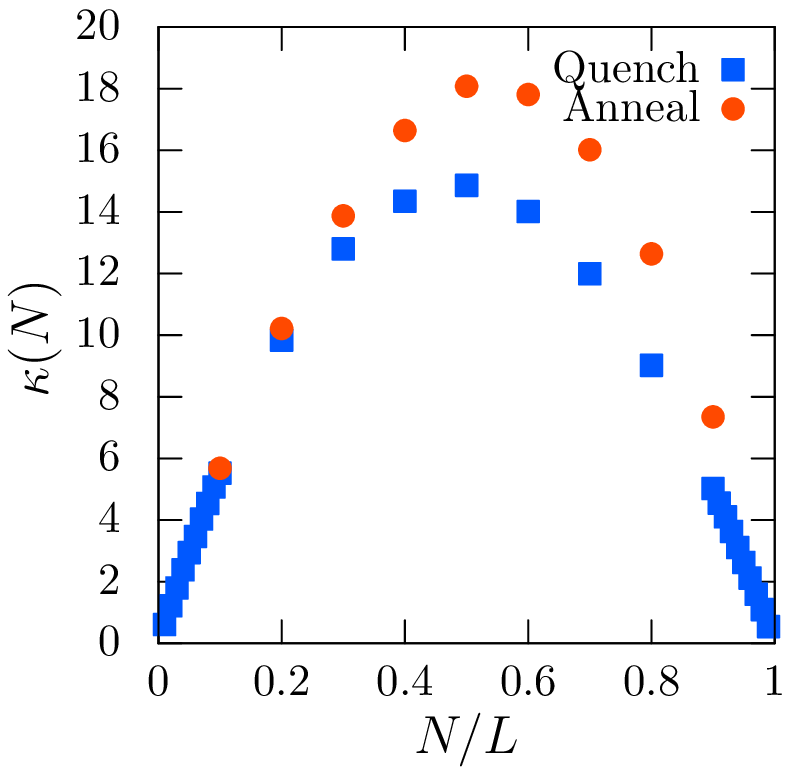}}
    \subfigure[\hspace{4cm}]{\includegraphics[width=4.2cm]{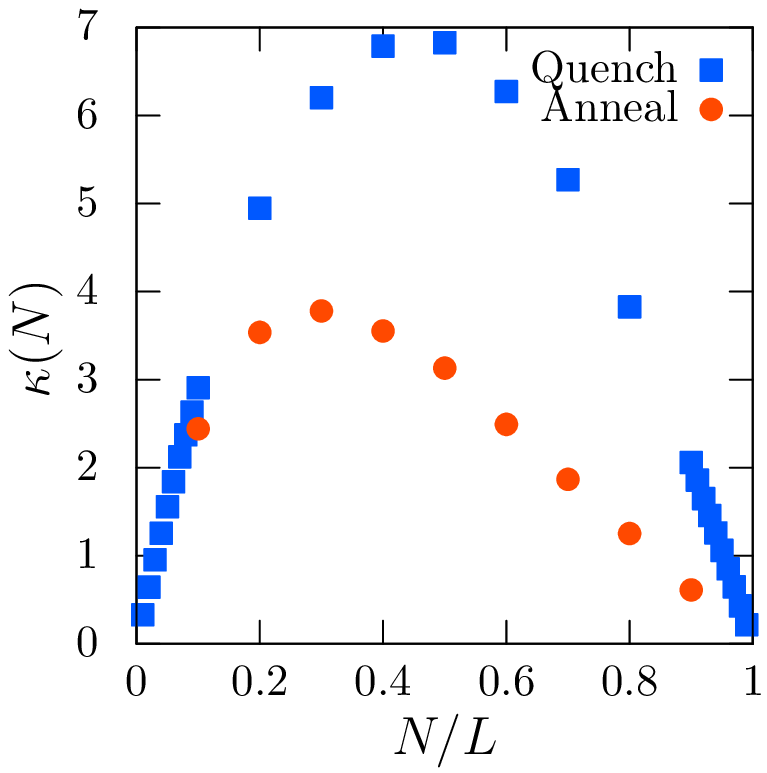}}
    \caption{
    Comparison between quenched and annealed results of the $\kappa(N)$-density relation ($L=100$, $T=1$, $\tau_c=1$).
    The distribution of the energy depth is the exponential distribution ($T/T_g=2.5$ for (a) and $T/T_g=1.5$ for (b)).
    Squares are the results of the numerical simulation of the dynamics of the SSEP on a quenched random energy landscape.
    Circles are the results of the numerical simulation of the dynamics of the SSEP on an annealed random energy landscape.
    }
    \label{fig: comparison between quenched and annealed}
\end{figure}

We compare the current fluctuations in the SSEP on the quenched and annealed random energy landscape.
The distinction between these two types of landscapes lies in whether the waiting time distribution depends on the site.
For the quenched random energy landscape, the waiting time distribution depends on the site,
while for the annealed random energy landscape, it does not.
For the ASEP, the current-density relation exhibits differences \cite{ConcannonBlythe,PhysRevE.107.L052103}.
In particular, for the quenched random energy landscape, the relation becomes flat around $N/L=1/2$.
In contrast, for the annealed random energy landscape, the relation becomes asymmetric around $N/L=1/2$
when the variance of waiting time diverges.
To distinguish between the cases where the variance of the waiting time is finite and where it diverges,
we consider that the distribution of the energy depth follows the exponential distribution.
For the annealed random energy landscape, the waiting time distribution follows the power-law distribution.
Figure~\ref{fig: comparison between quenched and annealed} shows that the relation between $\kappa(N)$ and density
becomes asymmetric around $N/L=1/2$.
When the variance is finite [Fig.~\ref{fig: comparison between quenched and annealed}(a)],
$\kappa(N)$ for the annealed random energy landscap is greater than that for the quenched random energy landscap.
Unlike the case of a finite variance, $\kappa(N)$ for the annealed random energy landscape is suppressed more than that for the quenched random energy landscape [Fig.~\ref{fig: comparison between quenched and annealed}(b)].
The asymmetry observed for the annealed disorder arises because, similar to the ASEP~\cite{ConcannonBlythe}, the interactions with surrounding particles lead to the divergence of the average waiting time.
Consequently, $\kappa(N)$ is surpressed significantly.

We discuss how to control the current fluctuations.
In the homogeneous environment, we can control the current fluctuations by adjusting the particle density.
On the other hand, in the heterogeneous environment, we can control the current fluctuations by regulating not only the particle density
but also the shape of the energy landscape.
For example, in the drug delivery device with uniform radii pores, the radii of the pores play a crucial role in controlling the drug release rate \cite{Yang:2010aa}.
Extending this concept, we propose that controlling current fluctuations in heterogeneous environments could be applied to drug delivery devices with non-uniform pore radii, where the shape of the energy landscape corresponds to the variation in pore radii.
Additionally, the understanding of controlling current fluctuations in heterogeneous environments may provide insights into biological transport phenomena, such as DNA-binding proteins~\cite{GraneliGreeneRobertsonYeykal,AustinCoxWang,Li:2009aa} and biological channels~\cite{AkimotoHiraoYamamotoYasuiYasuoka,10.1063/1.2761897}.

\section{Conclusion}\label{sec: Conclusion}
In this paper, we have studied how the disorder affects the current fluctuations in the SSEP within a heterogeneous environment.
First, we have provided the exact result \eqref{Second moment of total integrated current for single defect system} for the second moment of the integrated current for the SSEP with a defect site.
The asymmetry in the dependence of the second moment of the integrated current on the density around $1/2$
arises from the differing dynamics of holes and particles.
Surprisingly, in the low-density range, the second moment of the integrated current exceeds that of the homogeneous system.
Building on the insights gained from the SSEP with a defect site,
we have derived the approximate expression \eqref{Second moment of total integrated current for QTM} for the second moment of the integrated current for the SSEP on a quenched random energy landscape.
The behavior of the second moment of the integrated current is greatly influenced by the energy landscape.
Consequently, the presence of many-body effects induces sample-to-sample fluctuations in the fluctuation of the integrated current,
even when single-particle dynamics do not exhibit such fluctuations.
Our findings help control both current fluctuations and particle diffusivity by regulating the particle density.
This approach to controlling particle behavior has potential applications in drug delivery systems and biological transport phenomena.

\begin{acknowledgments}
    This work was supported by JST SPRING, Grant Number JPMJSP2151.
    T.A. was supported by JSPS Grant-in-Aid for Scientific Research (No. C JP21K033920).
\end{acknowledgments}

\appendix
\section{Derivation of Eq.~\eqref{second moment of total integrated current derivation}}
\label{sec: appendix current fluctuation}
Here, we derive the second moment of the total integrated current \eqref{second moment of total integrated current derivation}.
The time differential of the second moment of the total integrated current follows
\begin{widetext}
    \begin{equation}
        \begin{split}
            \frac{d\Braket{Q(t)^2}_N}{dt}&=\sum_{i=1}^L\Braket{\left[(Q(t)+1)^2-Q(t)^2\right]\left[\frac{1}{2\tau_i}\eta_i(1-\eta_{i+1})\right]}_N
            +\sum_{i=1}^L\Braket{\left[(Q(t)-1)^2-Q(t)^2\right]\left[\frac{1}{2\tau_i}\eta_i(1-\eta_{i-1})\right]}_N\\
            &=\sum_{i=1}^L\frac{1}{\tau_i}\Braket{\eta_i(1-\eta_{i+1})}_N
            +\sum_{i=1}^L\left(\frac{1}{\tau_i}-\frac{1}{\tau_{i+1}}\right)\Braket{Q(t)\eta_i\eta_{i+1}}_N,
        \end{split}
    \end{equation}
    where we use the detailed balance relation, $\Braket{\eta_i(1-\eta_{i+1})}_N/\tau_i=\Braket{\eta_{i+1}(1-\eta_i)}_N/\tau_{i+1}$.
    Therefore, for the SSEP with a defect site, we have
    \begin{equation}
        \frac{d\Braket{Q(t)^2}_N}{dt}=\sum_{i=1}^L\frac{1}{\tau_i}\Braket{\eta_i(1-\eta_{i+1})}_N
        +\left(1-\frac{1}{\mu}\right)\left[\Braket{Q(t)\eta_1\eta_2}_N-\Braket{Q(t)\eta_L\eta_1}_N\right].
    \end{equation}
\end{widetext}

\section{Time dependence of correlation between total integrated current and occupation number}
\label{sec: appendix correlation}
Here, we derive that the correlation between the total integrated current
and the occupation number does not depend on the time in the long-time limit.
Let $P_t(C,Q)$ be the probability that the system is in configuration $C$ at time $t$
and that $Q(t)=Q$.
The master equation of $P_t(C,Q)$ is
\begin{equation}
    \begin{split}
        \frac{dP_t(C,Q)}{dt}
        =&\sum_{C'}W_{+1}(C',C)P_t(C',Q-1)\\
        &+\sum_{C'}W_{-1}(C',C)P_t(C',Q+1)\\
        &-\sum_{C'}W(C,C')P_t(C,Q),
    \end{split}
\end{equation}
where $W_{\pm1}(C',C)$ is the rate that the configuration transitions from $C'$ to $C$ and that the total integrated current increases by $\pm1$,
and $W(C',C)=W_{+1}(C',C)+W_{-1}(C',C)$.
We introduce $P_t(C)$ and $q_t(C)$ defined by
\begin{align}
    &P_t(C)\equiv\sum_{Q}P_t(C,Q),\\
    &q_t(C)\equiv\sum_{Q}QP_t(C,Q).
\end{align}
$P_t(C)$ is the probability that the configuration of particles is $C$ at time $t$.
$q_t(C)/P_t(C)$ is the average of the total integrated current that the configuration of particles is $C$ at time $t$.
Notably, $q_t(C)$ is related to the correlation between the total current and the occupation number:
\begin{equation}
    \Braket{Q(t)\eta_i}_N=\sum_{C}\eta_iq_t(C).
\end{equation}
The time evolutions of $P_t(C)$ and $q_t(C)$ are Eq.~\eqref{master equation} and
\begin{equation}
    \begin{split}
        \frac{d q_t(C)}{dt}=&\sum_{C'}\left[W(C',C)q_t(C')-W(C,C')q_t(C)\right]\\
        &+\sum_{C'}\left[W_{+1}(C',C)-W_{-1}(C',C)\right]P_t(C'),
    \end{split}
    \label{appendix: time evolution of q}
\end{equation}
respectively.
For $t\rightarrow\infty$, the system reaches the steady state, and 
\begin{equation}
    P_t(C)\rightarrow P(C),\quad q_t(C)\rightarrow u(C)t+r(C),
    \label{appendix: q}
\end{equation}
respectively.
Substituting Eq.~\eqref{appendix: q} into Eq.~\eqref{appendix: time evolution of q}, we have
\begin{equation}
    \sum_{C'}\left[W(C',C)u(C')-W(C,C')u(C)\right]=0,
    \label{appendix: u}
\end{equation}
\begin{equation}
    \begin{split}
        &\sum_{C'}\left[W(C',C)r(C')-W(C,C')r(C)\right]\\
        &=u(C)-\sum_{C'}\left[W_{+1}(C',C)-W_{-1}(C',C)\right]P(C').
    \end{split}
    \label{appendix: r}
\end{equation}
Since Equation~\eqref{appendix: u} is the same as the equation for $P(C)$, $u(C)$ is given by
\begin{equation}
    u(C)=AP(C).
\end{equation}
Summing both sides of Eq.~\eqref{appendix: r} concerning $C$, we have
\begin{equation}
    \begin{split}
        \sum_{C}u(C)
        =&\sum_{C,C'}\left[W_{+1}(C',C)-W_{-1}(C',C)\right]P(C')\\
        =&0.
    \end{split}
\end{equation}
As a result, the parameter $A$ becomes $0$, i.e., $\lim_{t\rightarrow\infty}q_t(C)$ does not depend on the time.
Therefore, the correlation between the total current and the occupation number does not depend on the time in the long-time limit.

\bibliography{SEP_periodic}

\end{document}